\documentclass{elsart}
\usepackage{epsfig}

\usepackage{amssymb}
\begin{document}

\begin{frontmatter}



\title{Collective modes in strange and isospin asymmetric hadronic matter}
\thanks[label1]{This work was partially supported by the CONICET, Argentina.}

\author{R. Aguirre and A. L. De Paoli}

\address{Departamento de F\'{\i}sica, Fac. de Ciencias Exactas,\\
Universidad Nacional de La Plata.\\
C. C. 67 (1900) La Plata, Argentina.}

\begin{abstract}
We study the propagation of non-strange and strange meson modes in
hadronic matter considering both isospin and strangeness mixings
induced by quantum fluctuations in the medium. Baryons are
described using the Quark Meson Coupling model extended to include
interactions of strange quarks. In particular we evaluate the
dependence of the meson masses on the baryonic density, the
strangeness fraction and the isospin asymmetry of the medium. We
have found a considerable admixture of strangeness and isospin in
the $\sigma$-mode in the high density regime.
\end{abstract}

\end{frontmatter}
\oddsidemargin -1.1cm \small

 The study of the meson properties in a hot/dense hadronic
medium is at present an active field of research, since it is
related to fundamental aspects of the low energy regime of the
strong interactions such as the formation of a plasma of quarks
and gluons and the restoration of the chiral symmetry. A prominent
role is played by the lightest vector mesons, as they are closely
connected to the emission rate of dileptons in heavy ion collision
experiments. On the other hand the scalar mesons still keep
valuable information about the quark structure of hadrons as can
be seen, for example, in the dubious composition of the almost
degenerate
$a_0(980)-f_0(980)$ mesons.\\
There exist a profuse quantity of worthful studies of the
in-medium modification of meson properties, some of them are based
on pure quark models such as the Nambu Jona-Lasinio or use
descriptions in terms of hadrons only
\cite{CHIN,LIM,PAL,RHOME,RHODE,SIGOME,EPJA}. There were also some
efforts to reconcile both aspects by using hybrid
models such as the Quark Meson Coupling (QMC) \cite{ST1}.\\
A very interesting feature appearing in these treatments is the
mixing effect, which combines states of different isospin or
Lorentz components \cite{RHOME,RHODE,SIGOME}. This dynamical
violation of the lagrangian invariances is particularly important
in considering meson propagation, because
it opens new in-medium decaying channels.\\
The aim of this work is to study qualitatively the modification of
the meson properties in a dense medium as described by the
baryonic collective modes in the relativistic random phase
approximation (RRPA). In order to adjust to the conditions found
in the experience, as for instance in relativistic heavy ion
collisions, we consider matter with diverse strangeness fraction
as well as isospin asymmetry. We take into account the quark
structure of baryons by using the QMC model \cite{ST1,QMC,QMC2}.
In our treatment we regard all the
mixing polarizations allowed by the hadronic lagrangian proposed.\\
The evaluation of the RRPA with both isospin and strangeness
mixing effects is still lacking at present, and the qualitative
description of the lightest hadron masses in such a wide variety
of situations as found in the experience is also deficient.\\
The QMC model relates the in-medium properties of hadrons with
their quark structure.  Baryons are represented as non-overlapping
spherical bags containing three valence quarks, while the bag
radius changes dynamically with the medium density. The exchange
of  mesons coupled directly to the
confined quarks provide the baryon-baryon interaction.\\
To fit the model to our purposes we consider the strange
$f_0(980)$ and $\phi$ mesons as in Ref. \cite{QMC2}, and the
scalar iso-vector $a_0(980)$ together with the more commonly used
$\sigma$,
$\omega$, and $\rho$ mesons.\\
 The mean field approximation (MFA)
equation for a quark of flavor $q \; (q=u, d, s)$, of current mass
$m_q$ and $I_{3}^q$ third isospin component is
\begin{equation}
( i \gamma^{\mu} \partial_{\mu} - {m_q}^\ast - g_{\omega}^q
\gamma^0 \bar{\omega} - g_{\phi}^q \gamma^0 \bar{\phi} -
g_{\rho}^q I_{3}^q\; \gamma^0 \bar{\rho} ) \Psi^q = 0,
\label{QMCEQ}
\end{equation}
where we have used
\begin{eqnarray}
{m_q}^\ast&=&m_q - g_{\sigma}^q \,\bar{\sigma} - g_{\zeta}^q\,
\bar{\zeta} - g_{\delta}^q I_{3}^q\ \,\bar{\delta}. \label{QMASS}
\end{eqnarray}
In these Eqs. $\delta$ and $\zeta$ stand respectively for the
$a_0$ and $f_0$ fields, the upper bar indicates mean values. The
solution of Eq. (\ref{QMCEQ}) for a quark confined within a
spherical bag of radius $R_b$, representing a baryon of class $b$,
is well known and can be consulted for example in \cite{ST1,QMC}.
The baryon effective mass $M_b^\ast$ is given by
\begin{equation}
M_b^\ast=\frac{\sum_q N_q^b \Omega_{q b} - z_{0 b}}{R_b} +
\frac{4}{3} \pi B   {R_b}^3, \label{BAGMASS}
\end{equation}

where $N_q^b$ is the number of quarks of flavor $q$ inside the bag
and $\Omega_{q b} =[x_{q b}^2 +{(R_b {m_q}^\ast)}^2]^{1/2}$. The
eigenvalue $x_{q b}$ comes from the so called linear boundary
condition at the bag surface \cite{QMC}.
 The bag constant $B^{1/4}=210.86 MeV$ stands for non-perturbative vacuum
 contributions, it is adjusted to get a proton bag radius
 $R_p=0.6\, fm$ in vacuum. The zero-point motion parameters $z_{0 b}$
are fixed to reproduce the baryon spectrum at zero density. In our
calculations we have taken $m_u=m_d=5 MeV$ and $m_s=150 MeV$
for the current quark masses. \\
The equilibrium condition for the bag radius is \cite{OUR}\\

\begin{equation}
- \frac {1} {4 \pi R_b^2} {\left( \frac {\partial{M_b}^\ast}
{\partial R_b} \right)} = \frac{1}{3 \pi^2} \sum_{b'}
\int_0^{k_{b'}} \frac{dk k^4} {\sqrt{{{M_{b'}}^\ast}^2+k^2}}.
\label{QMCc}
\end{equation}

The sum on the r.h.s runs over all the baryonic species considered. \\
We neglect the coupling between strange and non-strange quarks and
mesons. To determine the remaining couplings we assume pure vector
dominance together with $SU(3)$ flavor symmetry, which allows to
regain the $SU(6)$ relations of the quark model, i.e.
\begin{equation}
g_{\sigma,\omega}^{b} = N^b_{ns}\, g_{\sigma,\omega}^{u},
\;g_{\delta,\rho}^{b} = |I_{3}^b|\, g_{\delta,\rho}^{u},\;
g_{\zeta,\phi}^{b} = N^b_{s} \, g_{\zeta,\phi}^{s}, \label{GC}
\end{equation}
 where $N^b_{ns}$ ($N^b_{s}$) is the non-strange
(strange) quark number inside the baryon $b$, with third isospin
component $I_{3}^b$, and in addition we have
$g_{\phi}^s=-\sqrt{2}\,g_{\omega}^u$. Furthermore as the $SU(2)$
flavor in the non-strange sector is almost perfectly realized, we
take as identical the coupling of quarks $u$ and $d$ with every
meson.\\
Therefore we have five independent couplings $g_{\sigma}^{u}$,
$g_{\omega}^{u}$, $g_{\delta}^{u}$, $g_{\rho}^{u}$, and
$g_{\zeta}^{s}$, we fix their numerical values by fitting
symmetric nuclear matter properties at saturation, namely baryonic
density $n_0 =0.15 fm^{-3}$, binding energy $E_b =-16 MeV$ and
symmetry energy $a_s =35 MeV$. The last condition gives a
functional relationship between $g_{\delta}^{u}$ and
$g_{\rho}^{u}$, we have chosen an arbitrary point on this curve.
To fix $g_{\zeta}^{s}$ we follow references \cite{PAL,QMC2,GAL}
assuming a potential depth $U_{\Xi}=40 MeV$ for the $\Xi$ hyperon
embedded in symmetrical $\Xi$ matter at baryonic density
$n_{\Xi}=n_0$. Thus we obtain $g_{\sigma}^{u,d}=5.99$,
$g_{\omega}^{u,d}=3.00$, $g_{\delta}^{u,d}=1.83$,
$g_{\rho}^{u,d}=4.50$, and
$g_{\zeta}^{s}=4.48$.\\

The self-consistent equations for the meson mean fields are of the
form

\begin{equation}
\bar{\chi}=-\frac{1}{m_\chi^2}\sum_b Y_\chi^b \, n_\chi^b,
\end{equation}
with $Y_\chi^b=\partial{M_b}^\ast/\partial\bar{\chi}$,
$n_\chi^b=n_s^b$ if the generic meson $\chi$ is a Lorentz scalar
field, or $Y_\chi^b= -g_\chi^\omega,-g_\chi^\phi,-g_\chi^\rho
I_{3}^b$, $n_\chi^b=n_b$ for the vector case. The scalar and
baryon densities
\[
n_s^b =\frac{1} {\pi^2} {M_ b}^\ast \int_0^{k_{F b}}
dk \frac{k^2} { \sqrt{{M_b^\ast}^2+k^2}}, \;\;
n^b= \frac {{k_{F b}}^3} {3 \pi^2} \nonumber\\ 
\]
are functions of the Fermi momentum $k_{Fb}$. The meson quantum
fluctuations will be considered later. For the hadronic masses in
vacuum we have used the values $m_\sigma=550$, $m_{\zeta}=980$,
$m_\omega=783$, $m_\phi=1020$, $m_\delta=984$, $m_\rho=770$,
$M_p=M_n=938.92$, $M_\Lambda=1115.63$, and $M_\Xi=1318.11$,
expressed in MeV.

 The QMC model describes effective baryons
propagating in a homogeneous background of classical meson fields
as described above. Beyond this picture we assume a linear
coupling between the mesonic fluctuations and the baryons emerging
from the MFA, with coupling constants given in Eq. (\ref{GC}). The
full meson propagator can be approximated in a non-perturbative
approach by summing the ring diagrams to all orders, i.e. the
RRPA. This is a common procedure in nuclear physics, and can be
used to evaluate the density dependence of the meson masses among
other properties. The QMC has been used previously to evaluate the
vector meson masses, regarding bosons as bags on the same foot as
the baryonic ones, but neglecting the quark structure of the $\sigma$ meson \cite{ST1}.\\
The one-loop proper polarization insertion describing the
propagation between meson states $\alpha$ and $\beta$ is given by

\begin{eqnarray}
\Pi_{\alpha \beta}(q)&=& - i\sum_b \, g_{\alpha}^b\; g_{\beta}^b
\int \frac{d^4k}{(2 \pi)^4} Tr\left[ G_b(q)\; \Gamma_{\alpha}\;
G_b(q+k)\; \Gamma_{\beta} \right], \label{POLARIZATION}
\end{eqnarray}
with indexes $\alpha, \beta$  running over the whole set of mesons
and its internal degrees of freedom.
 Further, $G_b(p)$ stands for the baryon $b$ propagator in the MFA,
and the structure of $\Gamma_{\alpha}$ depends on the baryon-meson
vertex, namely $\Gamma_{\sigma, \zeta}=~1,
\Gamma_{\delta}=\mathbf{\tau}, \Gamma_{\omega, \phi}=\gamma^{\mu}
, \Gamma_{\rho}=\gamma^{\mu}\, \mathbf{\tau}$. Eq.
(\ref{POLARIZATION}) contains ultraviolet divergences coming from
the vacuum contribution which require an appropriate
regularization. We adopt the dimensional regularization procedure,
preserving that $\Pi_{\alpha \beta}(q)=\Pi_{\beta \alpha}(q)$, and
we shall discuss later the choice of the regularization points.
Explicit expressions for Eq. (\ref{POLARIZATION}) are
similar to that given in \cite{LIM,SIGOME}, for instance.\\
The formalism is best described within a generalized meson
propagator in a matrix representation of dimension equal to the
sum of the mesonic degrees of freedom. For example the free
(${\mathcal{P}}^0$), and full (${\mathcal{P}}$) generalized meson
propagators have respectively in its diagonal blocks the free and
dressed meson propagators of the $\sigma, \zeta, \delta, \omega,
\phi,$ and $\rho$ fields, and in the complementary spaces they
have zeros or finite mixing polarizations arising from Eq.
(\ref{POLARIZATION}),
 respectively. The corresponding Dyson-Schwinger equation in
matrix form can be used to solve for ${\mathcal{P}}(q)$:
\begin{eqnarray}
{\mathcal{P}}(q)= {\left[ {\mathcal{I}}-{\mathcal{P}}^{0}(q)\;
\Pi(q) \right]}^{-1} \; {\mathcal{P}}^{0}(q) \label{DIELEC}
\end{eqnarray}

The dielectric function $\epsilon(q)=\det \left[ {
\mathcal{I}}-{\mathcal{P}}^{0}(q)\; \Pi(q) \right]$ is defined so
that its roots coincide with the poles of  ${\mathcal{P}}(q)$. We
study the temporal regime $\vec{q}=\vec{0}$, where scalar and
vector contributions to the equation $\epsilon(q)=0$ become
independent; in addition the longitudinal and transversal branches
of the vector mesons become degenerate. Hence we have two separate
equations of the form:
\begin{eqnarray}
(D_\lambda^{-1}-\Pi_{\lambda \lambda})(D_\mu^{-1}-\Pi_{\mu
\mu})(D_\nu^{-1}-\Pi_{\nu \nu})- (D_\lambda^{-1}-\Pi_{\lambda
\lambda})\,\Pi_{\mu \nu}-(D_\nu^{-1}-\Pi_{\nu \nu})\,\Pi_{\lambda
\mu}&=0, \label{ZEROES}
\end{eqnarray}

where the indices $\lambda, \mu, \nu$ take definite meson labels,
namely $\lambda=\delta$, $\mu=\sigma$, and $\nu=\zeta$ for the
scalar case, and $\lambda=\rho$, $\mu=\omega$, and $\nu=\phi$ for
the vector case.

Furthermore $D_\alpha^{-1}=q^2-m_\alpha^2$ is the inverse free
propagator for the $\alpha$-meson. It must be noted that there is
no $\delta$-$\zeta$ neither $\rho$-$\phi$ mixing, and this causes
the splitting of the isovector mode in vacuum as well as in the
isospin symmetric dense medium. This splitting appears because the
$\sigma$-$\delta$ and $\omega$-$\rho$ polarizations become null
under these conditions, in view of the degeneracy
assumed for the baryonic iso-multiplets.\\
 In order to extract
finite values from the polarization we require at zero baryonic
density \cite{AGUIRRE}
\begin{eqnarray}
&& \Pi_{\alpha \beta}(q^2=R_{\alpha\beta}^2)=0, \nonumber
\\
&&\left(\partial \Pi_{\alpha \beta}(q)/\partial q^2
\right)_{q^2=R_{\alpha \beta}^2}= (1-L_{\alpha \beta}) \sum_b
g_{\alpha}^{' b}\, g_{\beta}^{' b}/(8 \pi^2) \nonumber
\end{eqnarray}
The last equation holds only for both $\alpha$ and $\beta$
corresponding to scalar mesons, where $g_{\alpha}^{' b}=
g_{\alpha}^{b}$ for the isoscalars $\alpha=\sigma , \zeta$, or
$g_{\alpha}^{' b}= g_{\alpha}^{b}\, I_3^b$ for the isovector
$\alpha=\delta$. The regularizing parameter is fixed at $L_{\alpha
\beta}=10$ \cite{EPJA,AGUIRRE}. Here $R_{\alpha \beta}=R_{\beta
\alpha}$ are a set of regularization points that we choose in
order to reproduce the physical meson masses at zero density. This
requirement determines unambiguously $R_{\delta \delta}=m_\delta$,
and $R_{\rho \rho}=m_\rho$, otherwise we obtain a range of points
$(R_{\sigma \sigma},R_{\sigma \zeta},R_{\zeta \zeta})$ and
$(R_{\omega \omega},R_{\omega \phi},R_{\phi \phi})$. The
regularization points for the mixing polarizations including one
iso-vector meson are not constrained, for the reasons mentioned in
the preceding paragraph. In our calculations we have used
$R_{\sigma \sigma}=m_\sigma$, $R_{\sigma \zeta}=R_{\sigma
\delta}=2.807$, $R_{\zeta \zeta}=4.654$, $R_{\omega \omega}=3.98$,
$R_{\omega \phi}=R_{\omega \rho}=6.09$, $R_{\phi \phi}=5.12$
expressed in $fm^{-1}$. We have numerically checked that the final
conclusions are insensitive to the precise choice of these
parameters. Instead, the choice of $L$ has noticeable effects on
the behavior of the $\sigma$ meson properties \cite{EPJA,AGUIRRE},
and a detailed discussion within this context will be given elsewhere
\cite{OUR2}.

To study the behavior of meson masses in a hadronic medium with
strange and isospin content we have considered two situations: a)
nucleons and $\Lambda$, b) nucleons, $\Lambda$ and $\Xi$ in
equilibrium against strong decay. In the last case the relative
abundance of $\Lambda$ and $\Xi$ is determined by the relation
between their chemical potentials, a situation often treated in
the literature \cite{PAL,QMC2,GAL}. These two different hadronic
environments when considered at zero isospin asymmetry give
practically identical results for the meson masses. Therefore we
consider in the following only case (a).\\
 We take the total baryonic density
$n=(n^n+n^p+n^\Lambda)$ running up to $4 n_0$, guarantying that
baryonic bags do not overlap. For a fixed total baryonic density
$n$ we introduce the isospin asymmetry parameter $t=(n^n-n^p)/n$
and the strangeness fraction $S=n^\Lambda/n$, both $S$ and $t$
taking values in $(0,1)$.

\begin{figure}[t]\vspace{-3cm}
\begin{center}
\psfig{file=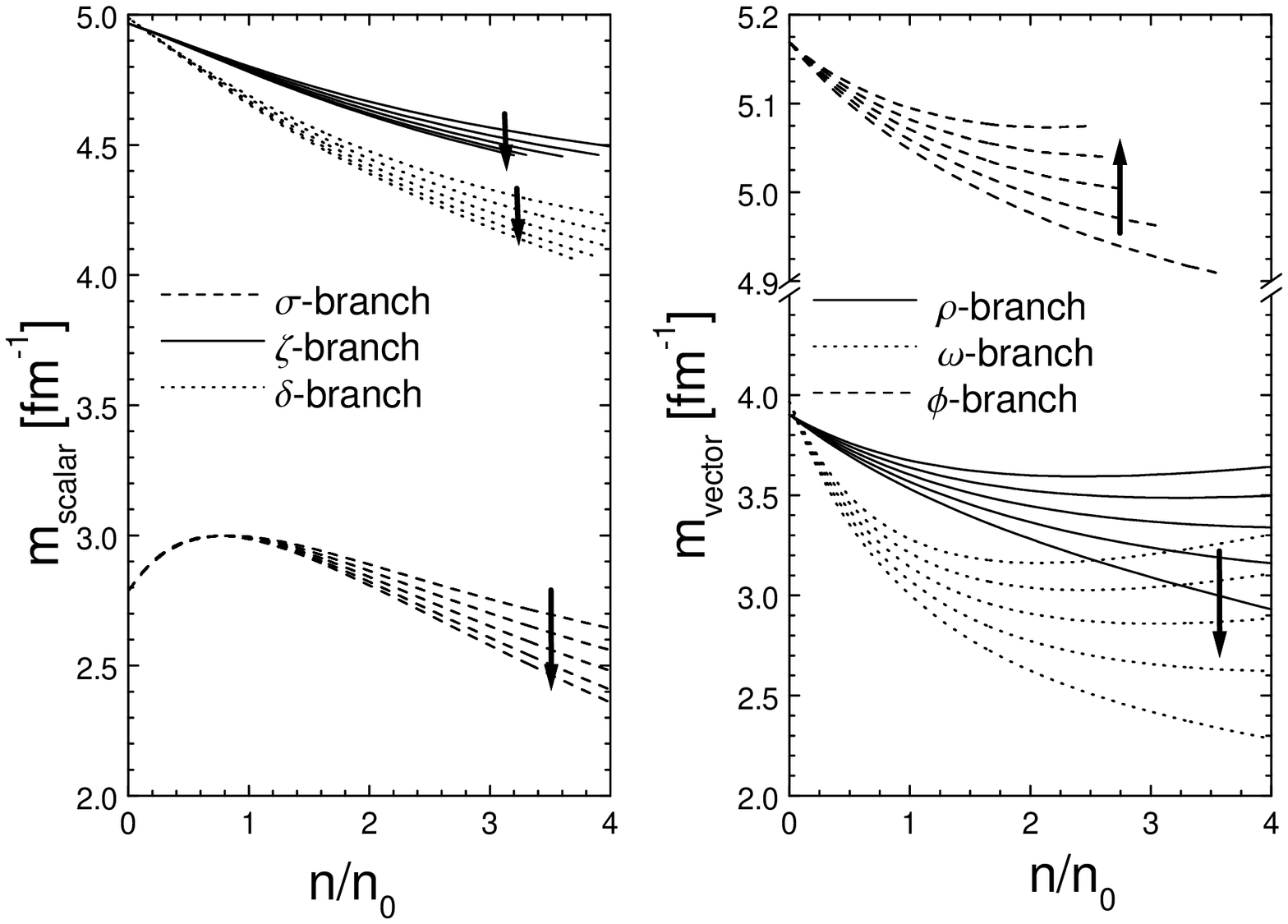,width=\textwidth}\end{center}\vspace{-5cm}
\caption{The effective meson masses as a function of the density
for $t=0$ and several strangeness fractions $S=0,\, 0.25,\, 0.5,\,
0,75,\, 1$. The left (right) panel corresponds to the scalar
(vector) meson branches. The arrows indicate the direction of
growing $S$.} \label{ONE}
\end{figure}

\begin{figure}[t] \vspace{-3cm}
\begin{center}
\psfig{file=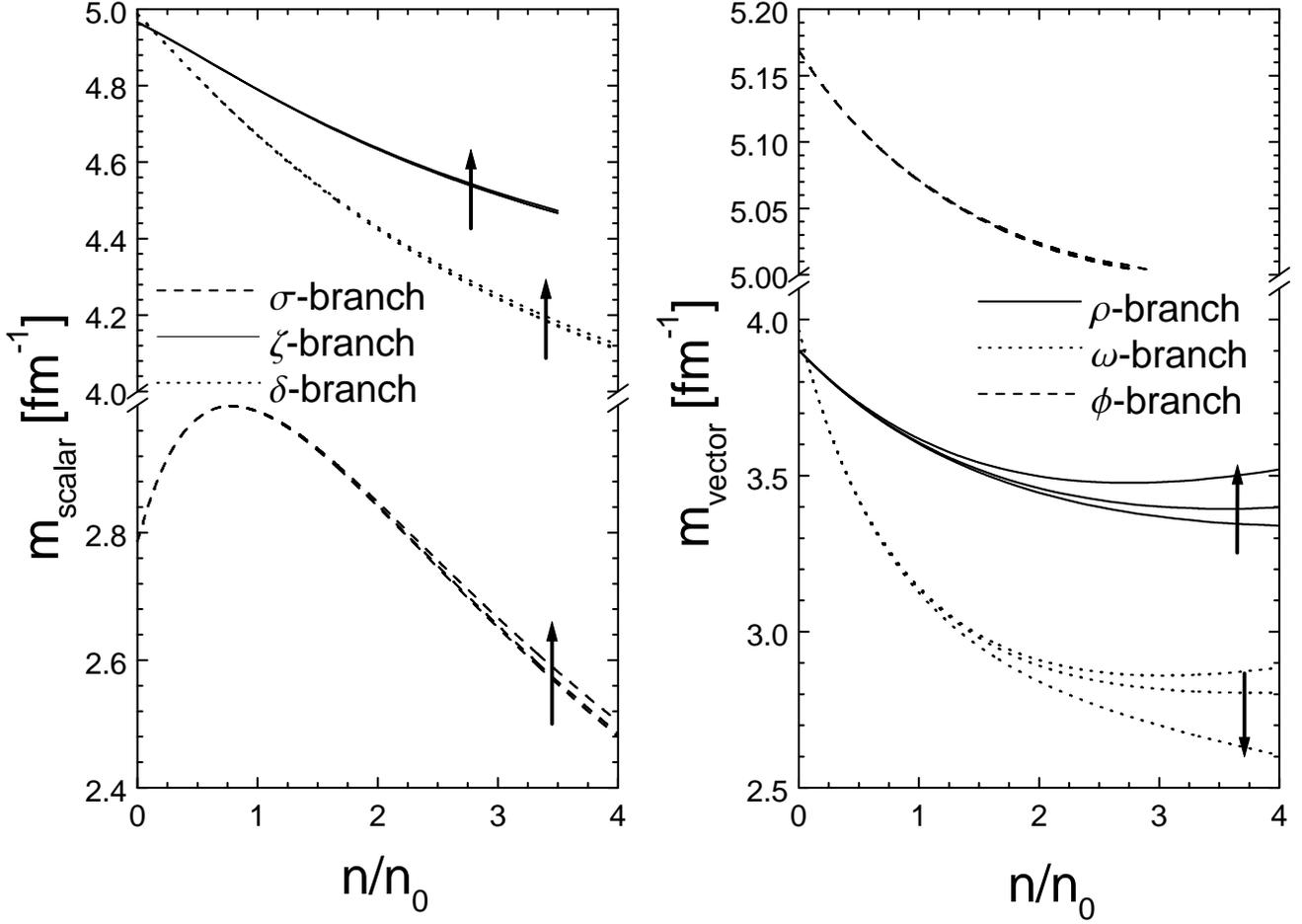,width=\textwidth}\end{center}
 \vspace{-5cm}
\caption{The effective meson masses as a function of the density
for $S=0.5$ and several isospin compositions $t=0,\, 0.5,\, 1$.
The left (right) panel corresponds to the scalar (vector) meson
branches. The arrows indicate the direction of growing $t$.}
\label{TWO}
\end{figure}

 In Fig. \ref{ONE} we plot the effective meson
masses computed as the zeros of Eq.(\ref{ZEROES}), for $t=0$ and
different strangeness fractions $S$. Although the meson quantum
numbers are blurred in-medium due to the mixing effect, we can
distinguish well defined masses branches continuously related to
its zero density value. Therefore we keep the label of the meson
which originates each branch.\\
 To explain the behavior of the meson masses shown in
this figure one must take into account that the polarization
insertion is composed of two additive terms of opposite sign,
namely the vacuum contribution $\Pi^{(F)}$ and the density
dependent part $\Pi^{(D)}$ which explicitly depends on the Fermi
momenta of the baryons. Thus, neglecting mixing corrections one
has $m_k^{\ast\, 2} \simeq m_k^2+\Pi^{(F)}_{kk}+\Pi^{(D)}_{kk}$,
$k= \zeta, \delta, \omega, \phi$ and $\rho$. At low densities ($n
\lesssim n_0$) vacuum fluctuations are dominant and tend to lessen
the meson effective masses. However the density dependent part of
the polarizations contributes in the opposite direction,
attenuating or even reverting this trend at higher densities. The
only exception is given by $m^\ast_\sigma$ for which $\Pi^{(F)}$
raises at low densities due to the relatively small value of
$m_\sigma$.\\
Within this approach the dependence on $S$ for a fixed baryonic
density can be explained by observing that $\Pi^{(F)}$ is a
negative decreasing function of $S$. On the other hand $\Pi^{(D)}$
decreases (increases) when $S$ increases for the non-strange
(strange) mesons, due to the depletion (population) of the baryons
to which they couple. Furthermore the variation of $\Pi^{(D)}$ is
stronger for the vector than for the scalar mesons. For instance,
as $S$ grows the total polarization of the $\phi$ branch increases
due to the predominance of $\Pi_{\phi \phi}^{(D)}$ over $\Pi_{\phi
\phi}^{(F)}$, therefore $m_\phi^\ast$  is enhanced. On the other
hand for the $\zeta$ branch, the growing of $\Pi^{(D)}$ is much
weaker and $\Pi^{(F)}$ determines the decreasing behavior of
$m_\zeta^\ast$ with $S$. The numerical results shown below provide
the justification for disregarding in first approach the mixing
effects for the strange mesons. This reasoning does not hold for
the $\sigma$ branch because it carries a strong mixing, leading to
a more involved behavior, but the final result
is a decreasing dependence with $S$.\\
On the other side, as can be appreciated from this figure, the
$\omega - \rho$ mass difference for baryon densities around $2
n_0$ is close to the pion mass, increasing the probability of the
$\rho \rightarrow \omega + \pi$ decay. This fact in turn
contributes to enhance the $\rho$ meson width, in agreement with
several theoretical predictions as well as with phenomenological
observations.\\
We can compare Fig. \ref{ONE} with the results shown in Ref.
\cite{PAL} (Figs. 3 and 4 for $q=1 MeV$) for the $\sigma, \omega$
and $\zeta$ mesons. It can be appreciated that the overall
behavior with $S$ agrees with our calculations within the
considered range of densities.\\
The influence of the isospin asymmetry $t$ over the meson modes at
constant strangeness $S=0.5$ is shown in Fig. \ref{TWO}. The
qualitative description does not change substantially for other
values of $S$. The strange meson masses are almost independent of
$t$, as they are coupled only to the iso-singlet $\Lambda$. On the
left hand panel, the $\sigma$-meson mass shows a slight
enhancement with growing $t$, this effect turns to be appreciable
for $n> 2 n_0$. On the opposite panel of this figure, the
isovector $\rho$-meson mass also exhibits a small enhancement with
$t$ for $n> 1.5 n_0$. Interestingly the isoscalar $\omega$-meson
mass presents the inverse trend when $t$ increases, therefore we
conclude that for a fixed strangeness content, the splitting
between the $\omega$ and $\rho$ masses increases with the isospin
asymmetry. These results can be compared, for example, with those
of Ref. \cite{RHOME} where the same behavior of $m^\ast_{\omega,
\rho}$ with $t$ is found.

A measure of the mixing between two physical meson states $\alpha$
and $\beta$ can be given by the mixing angle $\theta_{\alpha
\beta}$, defined by \cite{RHODE}
\begin{equation}
\tan\, 2 \theta_{\alpha \beta}=\frac{2 \Pi_{\alpha
\beta}}{m_\alpha^2+\Pi_{\alpha \alpha}-m_\beta^2-\Pi_{\beta
\beta}}. \label{MIXANGLE}
\end{equation}
Since our aim is to obtain a qualitative estimation of how much
isospin and strangeness remain good quantum numbers for the
collective modes, we do not include the width of the original
mesons in Eq. (\ref{MIXANGLE}) in consistency with the RRPA.
However quantitative calculations must include a imaginary
contribution to the polarizations \cite{RHOME,BRONIOWSKI}.

\begin{figure}[t]\vspace{-3cm}
\begin{center}
\psfig{file=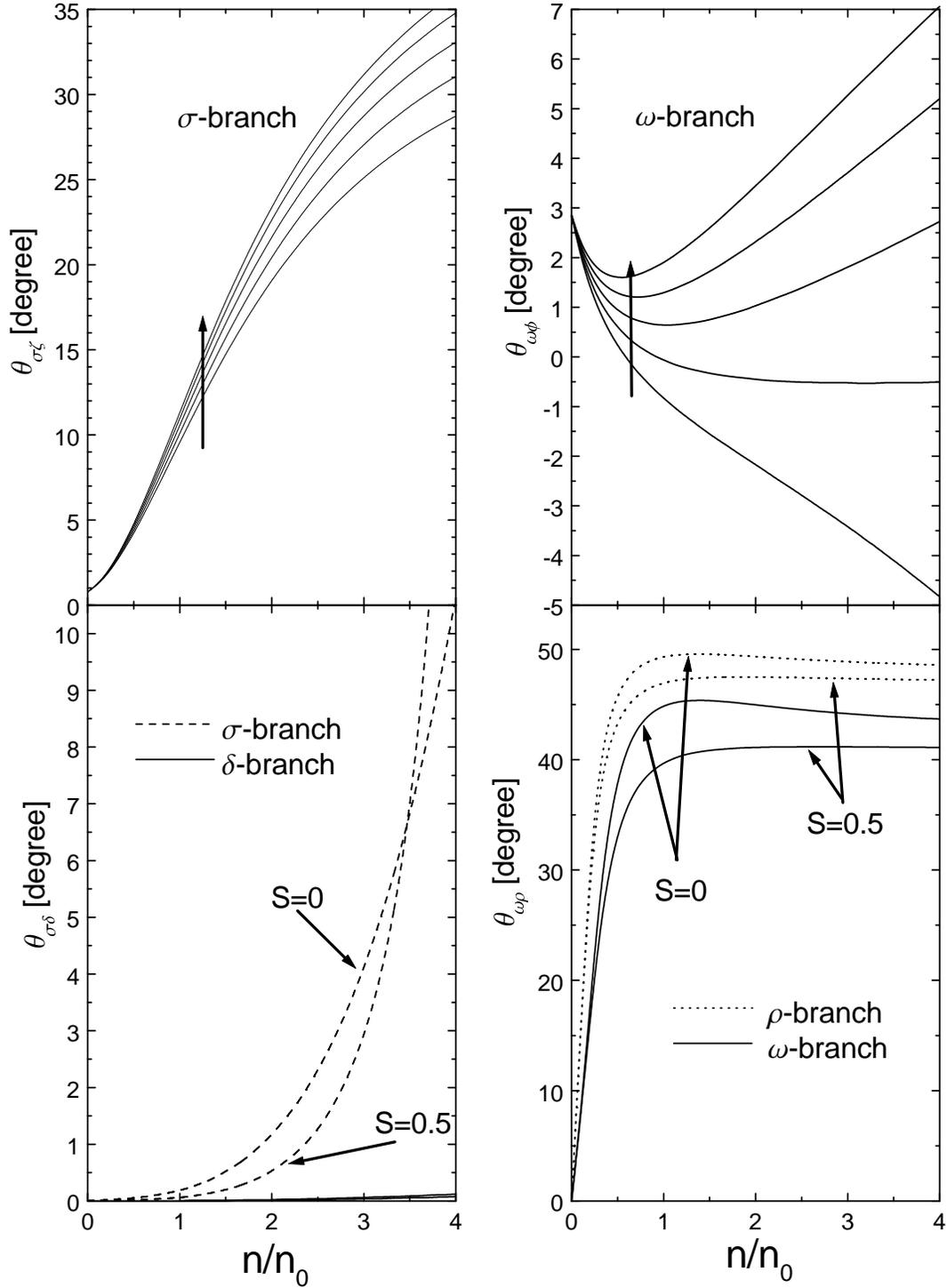,width=0.85\textwidth}\end{center}
\vspace{-1cm} \caption{The mixing angles defined by Eq.
(\ref{MIXANGLE}) as a function of the density. The upper panels
show strangeness mixing for $t=0$ and several strangeness
fractions $S=0,\, 0.25,\, 0.5,\, 0,75,\, 1$. The upper left (upper
right) panel corresponds to the scalar $\sigma$-$\zeta$ (vector
$\omega$-$\phi$) mix evaluated at the $\sigma$ ($\omega$) meson
branch. The arrows indicate the direction of growing $S$. Lower
panels show the isospin mixing angles for $t=0.5$ and the
strangeness fractions $S=0,\, 0.5$. The lower left (lower right)
panel corresponds to the scalar $\sigma$-$\delta$ (vector
$\omega$-$\rho$) mix evaluated at both the $\sigma$ and $\delta$
($\omega$ and $\rho$) meson branches.} \label{THREE}
\end{figure}

For $t=0$ the isoscalar-isovector mixing vanishes, remaining the
isoscalar strange-non-strange mixing only. We have evaluated
Eq.(\ref{MIXANGLE}) for $\alpha=\sigma$ ($\omega$), $\beta=\zeta$
($\phi$) at the pole $p_0=m_\sigma^\ast$ ($m_\omega^\ast$) in
terms of the baryonic density for various strangeness fractions
$S$. The result is plotted on the upper left (upper right) panel
of Fig. \ref{THREE}; as expected the mixing effect increases with
$n$ and $S$. We can observe that $\theta_{\sigma
\zeta}(q_0=m_\sigma^\ast)$ becomes appreciable at $n \gtrsim n_0$,
and therefore the mixing effects in the $\sigma$ channel may not
be neglected even at relatively low densities. In the case of
$\theta_{\omega \phi}(q_0=m_\omega^\ast)$ it turns to be
significative only for high densities ($n \gtrsim 3 n_0$).  With
respect to $\theta_{\sigma \zeta}(q_0=m_\zeta^\ast)$ and
$\theta_{\omega \phi}(q_0=m_\phi^\ast)$ (not shown here), their
amplitudes remain below $3.5^o$ for all the densities studied.\\
Because the strangeness mixing remains small for all but the
$\sigma$ meson, we can estimate the isoscalar-isovector mixing for
$t\neq 0$ by using Eq.(\ref{MIXANGLE}) also. In the lower left
panel of Fig. \ref{THREE} it can be seen that scalar
$\sigma-\delta$ mesons mix only appreciably in the $\sigma$ branch
at densities $n \gtrsim 3.5 n_0$. Instead, non-strange
$\omega-\rho$ vector mesons exhibit a noticeable isospin mixing of
approximately equal magnitude in both branches, as shown in the
lower right panel of this figure. However, inclusion of the
$\omega$ and $\rho$ widths is expected to decrease these values
\cite{RHOME,BRONIOWSKI}. By raising the strangeness fraction a
sizeable decrease of the mixing angle is obtained for both scalar
and vector mesons, in the considered range of $n$.\\
We conclude that the in-medium propagation of the $\sigma$-channel
concentrates non-negligible admixtures of isospin as well as
strangeness in the high density regime. The vector $\omega$ and
$\rho$ modes are affected by isospin but not by strangeness
mixing. On the other hand the $\delta, \zeta,$ and $\phi$ modes
propagate in almost pure states for all the densities studied.\\
The effects discussed for vector mesons could be observable for
instance, in the low mass dilepton production from heavy ion
collision. The in-medium modification of the vector mesons has
been successfully invoked to explain the gross features of the
dilepton spectra from the CERES collaboration \cite{LiKoBrown}, in
the frame of the Vector Meson Dominance model. A key role is
assigned to the in-medium dropping of the $\rho$-meson mass. In
order to get this result at the tree level, a direct coupling of
the $\sigma$ to the $\omega / \rho$ fields was proposed
\cite{LiKoBrown}. In our approach we obtain the same qualitative
behavior within the RPA, and also a description of the mixture of
states. Because of the smallness of the $\omega - \phi$ mixing
shown in Fig. \ref{THREE} and the fact that both mesons decay
largely after the freeze out, we do not expect a significative
modification in these channels. The $\omega - \rho$ mix is large
enough at almost all densities to have significative consequences.
Because the $\rho$ and $\omega$ mesons decay mainly into two and
three pions respectively, in the dense hadronic medium the
alternative mechanism $\rho \rightarrow \omega \rightarrow 3 \pi$
is viable, having the effect of removing part of the strength
between the peaks corresponding to both mesons and transferring it
to the left of the $\rho$ peak in the dilepton spectra.

Some points deserve to be investigated within this model, for
example the meson coupling to Goldstone bosons, the fact that the
non-overlapping bag hypothesis is violated in the extreme density
regime used in astrophysical applications \cite{OUR}, and the
medium dependence of the scalar coupling constants which play a
significative role in chiral models \cite{RHO,CHANFRAY}.

\newpage

\end{document}